\documentclass{ws-ijmpcs}

\begin{document}

\markboth{Y. Ashitani, K.-I. Imura, \& Y. Takane}
{Perfectly Conducting Channel and Its Robustness
in Disordered Carbon Nanostructures}

%
\catchline{}{}{}{}{}
%

\title{PERFECTLY CONDUCTING CHANNEL AND ITS ROBUSTNESS
IN DISORDERED CARBON NANOSTRUCTURES}

\author{YUKI ASHITANI$^1$, KEN-ICHIRO IMURA$^{1,2}$, 
YOSITAKE TAKANE$^1$\footnote{Author to whom correspondence should be addressed.}}

\address{$^1$Department of Quantum Matter,
Graduate School of Advanced Sciences of Matter,\\
Hiroshima University, Higashihiroshima, Hiroshima, 739-8530, Japan;
$^2$Kavli Institute for Theoretical Physics, University of California, Santa Barbara, 
CA 93106, USA.}

\maketitle


\begin{abstract}
We report our recent numerical study on
the effects of dephasing on a perfectly conducting channel (PCC), 
its presence believed to be dominant in the transport characteristics of
a zigzag graphene nanoribbons (GNR) and of a metallic carbon nanotubes (CNT).
Our data confirms an earlier prediction that a PCC in GNR exhibits
a peculiar robustness against dephasing, in contrast to that of the CNT.
By studying the behavior of the conductance
as a function of the system's length we show that
dephasing destroys the PCC in CNT, whereas it stabilizes the PCC in GNR. 
Such opposing responses of the PCC against dephasing stem from
a different nature of the PCC in these systems.

\end{abstract}

\ccode{PACS numbers: 72.80.Vp, 73.20.Fz}

\section{Introduction}

In one-spatial dimension any weak disorder is believed to have
the potentiality of converting a good metal to an insulator,
as a consequence of the Anderson localization.\cite{imry}
However, the existence of two counter examples for this common belief
has been pointed out recently, both being found in a carbon nanostructure,
exhibiting a perfectly conducting channel (PCC).
The two examples are
i) the metallic carbon nanotube (CNT),\cite{ando1}\cdash\cite{suzuura}
and ii) the zigzag graphene nanoribbon (GNR) with edge modes of
partially flat dispersion.\cite{wakabayashi1}\cdash\cite{wakabayashi3}
A PCC is immune to backward scattering;
its existence allowing the conductance of the system to remain {\it finite}
even when its length $L$ becomes infinitely long, indicating
the {\it absence} of Anderson localization.
Note also that both CNT and GNR can be regarded as a derivative form of
an infinitely large graphene sheet
possessing two energy valleys around its Dirac points $K$ and $K'$.
Since scattering between these two valleys,
i.e., the inter-valley scattering,
usually destroys the perfectly conducting channel,
we focus on the case in which the system is subject to
only long-ranged scatterers, i.e., impurities whose potential range is
larger than or comparable to the size of the unit cell.

\begin{figure}[tpb]
\centerline{\psfig{file=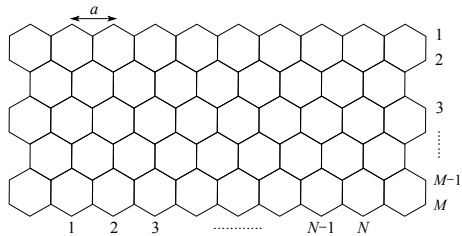,width=6.0cm}}
\vspace*{8pt}
\caption{Real space image of a GNR consisting of $M$ zigzag lines.
A CNT consisting of the same number of zigzag lines can be obtained by 
rolling up this GNR and linking each site of the first zigzag line
with its partner on the $M$th row.
\label{zigzag}}
\end{figure}

This paper highlights the behavior of such a PCC
believed to be existent in the carbon nanostructures.
Since a PCC appears within a quantum-mechanical framework,
one may think that it is fragile against a loss of the phase coherence
due to inter-electronic Coulombic interaction, electron-phonon coupling, etc.
This naive speculation, however, turns out to be not necessarily the case,
as we further elaborate the description of this phenomenon below.
We have performed extensive numerical study of such carbon-based
disordered quasi-one-dimensional systems
using the standard tight-binding representation of
the graphene's honeycomb lattice structure (see Fig.~\ref{zigzag}).
Our treatment of the dephasing follows that of Ref.~\refcite{suzuura}.

\section{Perfectly conducting channels in GNR and CNT}
\begin{figure}[btp]
\centerline{\psfig{file=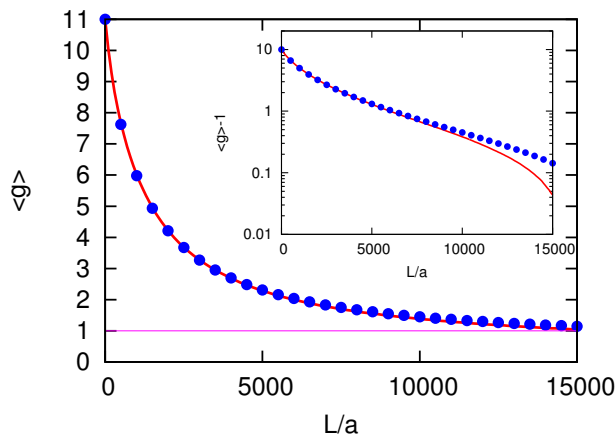,width=8.0cm}}
\vspace*{8pt}
\caption{Conductance of a disordered graphene nanoribbon:
a linear plot of the dimensionless conductance $\langle g \rangle$
(main panel), and a semi-log plot of $\langle g \rangle -1$ (inset)
as a function of the length $L$ of the disordered region
measured in units of the lattice constant $a$.
Solid lines (filled dots) corresponds to the case without (with) dephasing.
We set $M=30$ and $\epsilon_F/t = 0.579$ for which
the total number of conducting channels is $11$
(i.e., $g = 11$ at $L/a \rightarrow 0$).
Other parameters are $W/t = 0.13$, $p = 0.1$ and $L_{\phi}/a = 500$,
where $W$ measures the strength of each scatterer, and
$p$ is the probability that each site is occupied by a such scatterer.
The ensemble average is performed over $10^4$ samples with different
impurity configurations.
The magnitude of the error bar at $L/a = 15000$ is of order $10^{-3}$.
\label{plot-g}}
\end{figure}

In the case of GNR with zigzag edge boundaries,
the existence of a PCC is originated from its peculiar band structure.
Indeed, one can give it a simple interpretation based on
the appearance of partially flat-band edge modes.\cite{fujita}
Since these flat bands appear only in a part of the one-dimensional
Brillouin zone connecting the two valleys,
if one counts the number of conducting channels
of each propagating direction at a given Fermi energy,
there always exist an excess right-going channel in one valley
and an excess left-going channel in the other valley.
Let $N_{c}$ be the number of conducting channels in each valley
in the absence of the edge modes.
The above fact indicates that the number of right-going (left-going)
channels is $N_{c}$ ($N_{c}+1$) in one valley
and $N_{c}+1$ ($N_{c}$) in the other valley.
This imbalance leads to the appearance of one PCC which is robust against
disorder,\cite{barnes1,hirose} resulting in a noteworthy statement
on the scaling of the dimensionless conductance $g(L)$,
i.e., ``$g(L)$ scales naturally to a smaller value as the length $L$ of
the disordered region increases, but in the large-$L$ limit,
$g(L)$ remains to be a finite as $\lim_{L\rightarrow\infty}
g(L) = 1$.\cite{wakabayashi1}\cdash\cite{wakabayashi3}
Interested readers may refer to Ref.~\refcite{takane_incoherent}
and references therein for more detailed discussion
on the transport characteristics of such a system
with an imbalance in the number of right- and left-going channels.
In a recent paper,\cite{takane_incoherent} one of the authors has shown
that this PCC still survives even in the incoherent regime,
where information on the phase of the electronic wave function
is essentially lost.
This unexpected robustness of the PCC in GNRs in the presence of dephasing
(see also Fig. \ref{plot-g}) stems most certainly from the fact
that the imbalance in the number of conducting channels is not a consequence
of a particular symmetry
(cf. role of the so-called ``pseudo-time reversal'' symmetry
in the CNT case, see the discussion below);
it is simply guaranteed by the existence of
partially flat-band edge modes.

\begin{figure}[bth]
\centerline{\psfig{file=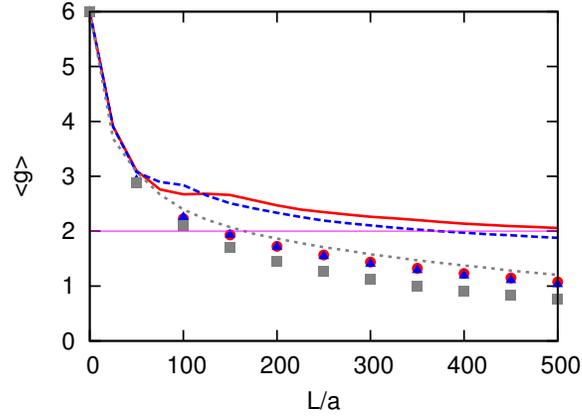,width=7.5cm}}
\vspace*{8pt}
\caption{Conductance of a disordered metallic carbon nanotube
as a function of the length $L$ of the disordered region
measured in units of the lattice constant $a$.
Solid; broken; dotted lines (filled circles; triangles; squares) correspond
to the case without (with) dephasing,
and of a different diameter of the nanotube:
$M=150$; $M=100$; $M=50$.
The three cases are also represented by different colors: red; blue; gray.
We set $\epsilon_{F}/t = 0.042$, $0.06309$, and $0.12623$, respectively
to the above three cases so that the initial value of $\langle g \rangle$
always takes the same value: $\langle g \rangle_{L \rightarrow 0} = 2 N_c = 6$.
Other parameters are set as
$W/t = 0.3$ and $p = 0.1$ and $L_{\phi}/a = 50$.
The ensemble average is performed over $5000$ samples.
The magnitude of the error bar at $L/a = 500$ is of order $10^{-3}$.
\label{plot-c}}
\end{figure}

In contrast to the case of GNR, the existence of a PCC in CNTs is
a much subtle issue. 
It is certainly essential that the system belongs to
the symplectic symmetry class, i.e., 
the total Hamiltonian of the system inclusive of the random potential
must, not only be time-reversal symmetric (TRS), but also fall on the case of
$\Theta^2=-1$ with $\Theta$ being the time-reversal operator.
This is typically the case with an effective spin-$1/2$ system
of a Dirac-type conic dispersion relation
though in this case TRS is not a real one (often dubbed as ``pseudo-TRS'').
This condition, therefore, will be safely satisfied in CNTs
under the influence of long-ranged potential disorder.
However, this condition alone turns out to be still not a sufficient one
for ensuring the existence of a PCC.
Much work on this subtlety,
associated with the parity of the number $N_c$ of the conducting channels
in each single Dirac cone,
has been pursued by Ando and co-workers in the context of studying
the transport characteristics of CNTs at a very early stage in the development
of this field.\cite{ando1}\cdash\cite{nakanishi}
To the best of our knowledge a clear statement on the condition for
the appearance of PCC, i.e., the idea that {\it both} of the following
two conditions:
i) appurtenance to the symplectic symmetric class,
ii) oddness of the number of conducting channels,
must be satisfied, has first appeared in Ref.~\refcite{suzuura}.
Notice that in the band structure of metallic CNTs,
only the single lowest gapless subband (of quasi-linear dispersion)
is non-degenerate,
whereas other quadratic subbands are all two-fold degenerate.
Therefore, wherever the Fermi level $\epsilon_F$ is,
the number of conducting channels in each propagating direction
in a given valley is necessarily odd.
This ensures the existence of at least one PCC per valley
(cf. Fig.~\ref{plot-c}).
Clearly, the dimensionless conductance $g$ ($=2 N_c$ in the clean limit;
here the factor 2 comes from the two valleys) decreases as disorder increases,
but remains finite due to the appearance of two PCCs.
This can be rephrased as follows:
``For a fixed strength of disorder, $g$ scales down to a smaller value
as the system becomes longer (as $L$ increases), but it approaches
asymptotically to an integral value, which is 2, in the long-$L$ limit''.
Such a behavior of the so-called ``symplectic-odd symmetry class'' has been
more profoundly elucidated
by the subsequent studies\cite{takane_DMPK}\cdash\cite{sakai}
in the context of the DMPK equation and the supersymmetric field
theory.\footnote{It seems fair to mention that a similar idea
but in a different context has already appeared
in a earlier work of Zirnbauer and co-workers.\cite{zirnbauer,mirlin}}
The existence of PCC in CNTs relies on the presence of pseudo-TRS.
Therefore, it could be fragile against any disturbances that might cause
breaking of the pseudo-TRS,
e.g., against trigonal warping of the Dirac cone.\cite{akimoto}
It is, therefore, natural to presume that PCC might be
fragile against dephasing.\cite{suzuura}
In this paper, we have extended this consideration
on the role of dephasing in the robustness of PCC in CNTs,
primarily for the comparison with the GNR case,
but with much care to the dependence on the circumference $R$
of the nanotube.\footnote{The larger the circumference $R$ is, 
the more closely are the subbands spaced.
Also, the further one goes away from the Dirac point, 
the stronger the trigonal warping becomes in the spectrum of a CNT.
Combining these two observations, one immediately realizes that
for a fixed value of $\epsilon_F$ and a given number of $N_c$,
the warping effects become stronger with decreasing $R$,
leading to stronger pseudo-TRS breaking.}

\section{Sketch of the numerical analysis and its implications}

Let us consider again the case of a GNR with $M$ zigzag lines
as shown in Fig.\ref{zigzag}.
The electronic states in this nanostructure
is described by a tight-binding Hamiltonian,
\begin{equation}
\label{ham}
  H = - \sum_{i,j} \gamma_{i,j}|i\rangle \langle j|
      + \sum_{i}V_{i}|i\rangle \langle i| ,
\end{equation}
where $|i\rangle$ and $V_{i}$ represent the localized electron state
and the impurity potential, respectively, on site $i$,
and $\gamma_{i,j}$ is the transfer integral between sites $i$ and $j$
with $\gamma_{i,j} = t$ if $i$ and $j$ are nearest neighbors
and $\gamma_{i,j} = 0$ otherwise.
We assume that the zigzag lines are infinitely long.
Instead, we distribute impurities (randomly) only in a finite region
(the disordered region, composed of $N$ columns)
of this infinitely long ribbon.
What we have been calling the ``system's length $L$'' so far is
now identified as the length $N$ of this disordered region, i.e., $L/a = N$.

In the actual computation, we have numerically estimated
the dimensionless conductance $g(L)$
using the Landauer formula and recursive Green's function method.
We assume that the potential profile of the scatterers is gaussian with
its characteristic range $d$ chosen to be $d/a = 1.5$,
a value large enough for avoiding the inter-valley scattering.
We then let the amplitude of this gaussian random potential $w$
be uniformly (randomly) distributed within the range of $|w| \le W/2$.
As we mentioned earlier, the effects of dephasing has been taken account of
by the approach employed also in Ref.~\refcite{suzuura},
i.e., by separating the entire sample into several segments of
equal length $L_{\phi}$.\footnote{The rule of this game is the following:
each time the incident electron leaves a segment and enters the next
one, he loses his phase memory.
As for concrete implementation of this to realistic carbon nanostructures,
we refer interested readers to our forthcoming publication.}

Let us now look at Fig.~\ref{plot-g}.
The main panel shows a linear plot of $\langle g \rangle$,
indicating that $\langle g \rangle$ converges to unity
irrespective of the presence or absence of dephasing;
a clear signature of the appearance of a PCC.
This partly confirms numerically our earlier prediction
based on a Boltzmann equation approach, stating that 
``the PCC in a GNR is so robust that it may possibly survive
even into the incoherent regime.''\cite{takane_incoherent}
In our plots one can also observe that $\langle g \rangle$ in the presence
of dephasing is slightly larger than the case of no dephasing.
This feature is more clearly highlighted in the semilog plot
of $\langle g\rangle - 1$.
When $L/a \gtrsim 10^4$ (i.e., $L$ is very large),
the value of $\langle g\rangle - 1$ without dephasing scales away
from a quasi-linear (stable) behavior in the presence of dephasing.
This is probably due to 
residual inter-valley scattering.
Notice that here dephasing plays indeed the role of
{\it stabilizing the PCC} against weak inter-valley scattering.

Let us finally analyze our numerical data for CNTs (Fig. \ref{plot-c}).
We make a few remarks on our CNT data,
which show a number of contrasting features to the case of GNR.
First, the value of $\langle g \rangle$ is {\it smaller}
in the presence of dephasing than in the absence of dephasing,
which is consistent with the result of Ref.~\refcite{suzuura}.
This simply opposes the GNR case.
In some cases ($M = 50$ and $100$) $\langle g \rangle$ decreases
even below the ``protected'' value of 2 as $L/a$ increases.
As mentioned earlier, trigonal warping of the Dirac cone is omnipresent
whenever the Fermi level is away from the Dirac point,
and this can possibly come into play in the transport characteristics
of a CNT,\cite{akimoto} when its diameter or $M$ is not large enough.
This seemingly weak effect associated with the breaking of pseudo-TRS
is shown to give a destructive influence on the scaling behavior
of $\langle g \rangle$ in the large-$L/a$ limit.
Dephasing does not help.
These observations lead us to our second conclusion that
a stable existence of the two PCCs in a CNT is restricted
to the case of a very {\it large diameter}
and of a relatively {\it small doping}.

\section*{Acknowledgments}

This work was supported in part by a Grant-in-Aid for Scientific
Research (C) (No. 21540389)
from the Japan Society for the Promotion of Science,
and by the National Science Foundation under Grant No. NSF PHY05-51164.


\end{document}